# HR Lyrae (Nova Lyr 1919): from outburst to active quiescence

Jeremy Shears & Gary Poyner

Nova Lyrae was discovered at the Harvard College Observatory on 1919 December 6 at magnitude 6.5. We present a lightcurve for this nova based on published and archival observations. This was a classical fast nova, probably of type B. Decline times were $t_2$ = 31 or 47d, depending on the method used, and $t_3$ = 97d. The amplitude was at least 9.5 magnitudes. Based on our $t_2$ values, we estimate that the absolute magnitude at maximum was −6.9 or −7.2 (±1.1) and at minimum is +2.3 or 2.6 (±1.1). The star shows an active quiescence with brightness variations on a variety of timescales. Visual observations over a period of ten years also reveal long periods when the star was around 15.4v and others when it was close to 15.7v. Finally, we point out that some characteristics of the star are similar to those of recurrent novae and propose further monitoring of future activity.

## An introduction to novae

One of the most spectacular astronomical events is the occurrence of a nova. Such 'guest stars', whose sudden appearance can change the view of the heavens, have been observed since antiquity. During the last century more than a dozen novae were bright enough to be visible to the unaided eye. The brightest of them all was Nova Aquilae 1918 (now known as V603 Aql), which for a time outshone all the other stars in the night sky apart from Sirius. The excitement and wonder associated with the appearance of Nova Aql was described by the American amateur astronomer, Leslie Peltier, in his book *Starlight Nights*. Earlier the same day he had observed an eclipse of the Sun with his small telescope:

> 'When darkness came that evening I clamped my spyglass to the grindstone mount which was still standing at the station underneath the walnut tree. I hoisted it up on my shoulder and carried it out to the middle of the front yard and stood it where I would have a clear view of the variable stars in the southeastern sky. That was the night that I forgot all about telescopes and variables for as I turned and looked up at the sky, right there in front of me − squarely in the centre of the Milky Way − was a bright and blazing star!'[1]

Novae are a class of cataclysmic variable star, which are known to be interacting binaries where a cool secondary star loses mass to a white dwarf primary.[2] Material from the secondary falls through the inner Lagrangian point and because it carries substantial angular momentum, it does not settle on the primary immediately, but forms an accretion disc around the primary. Material flowing through the accretion disk accumulates on the surface of the white dwarf and eventually causes a runaway thermonuclear reaction. The ensuing explosion causes the system to increase in brightness dramatically, blowing the outer layers of the white dwarf away into space as an expanding gas shell. With time, the gas cools and the once bright star begins to fade: the outburst is over.

Novae can be classified by the time it takes for the brightness to decline by three magnitudes. Fast novae decline in less than 100 days; by contrast slow novae take more than 100 days. Fast and slow novae are generally referred to as classical novae. A further class of novae is the 'recurrent novae', which undergo two or more outbursts, usually separated by many years.

## Discovery of Nova Lyr 1919 and its initial decline

Nova Lyr was discovered on 1919 December 6 by Miss Mackie during the course of her systematic search for novae in the Milky Way on photographs taken at the Harvard College Observatory (HCO).[3] The discovery announcement in the HCO *Bulletin* states:

> 'Miss Cannon finds that [the star] has the characteristic spectrum of the early nova type. Between Dec 4 and 6 it rose rapidly from the 16th magnitude or fainter, to a maximum of about 6.5. Since that time it has undergone marked fluctuations in brightness. Its present magnitude is 8.5 (1920 Jan 6).'

Conditions could not have been more different from the discovery of Nova Aql the previous year. The new nova was not especially bright and December is an unfortunate time for any object to be discovered in Lyra, due to the position of the constellation low in the northern sky at this time of year. Furthermore, by the time the news was relayed to Europe, it was already 1920 January and the observing conditions were even less favourable. As a consequence rather few observations of the nova exist for the first few weeks after its appearance.

In order to examine the early phase of the nova's brightness thoroughly, we have compiled a lightcurve based on published observations (Figure 1), as well as those in the AAVSO and AFOEV (French variable star association) databases. We believe this is the most detailed light curve of this phase of the outburst presented to date. We have so far found no record of observations of the nova in the BAAVSS archives during this critical time. However, we did find some observations in the literature made by long-time director of the VSS, Felix de Roy,[8] during 1920 February and March, and by W. H. Steavenson in later years. Subsequently, the nova was assigned the name HR Lyr[9] and we shall use this designation in the rest of this paper. HR Lyr is located at RA 18h 53m 25.0s, Dec +29° 13m 37s (J2000.0).

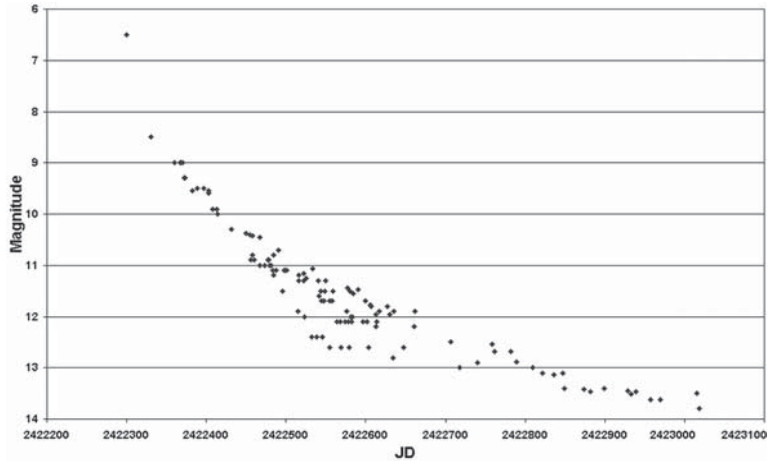

**Figure 1.** Lightcurve of HR Lyr from 1919 Dec 4 to 1921 Nov 24. Data are from references 3 to 8 and the AAVSO and AFOEV databases.

## Decline parameters

Two parameters are commonly used to described the decline phase of a nova's light curve: $t_2$ and $t_3$. These are the times taken for the nova to drop by 2 and 3 magnitudes respectively below maximum. Given the paucity of data surrounding the detection of the nova, it is difficult to be certain of the time and magnitude at maximum: we know from the HCO *Bulletin* that the nova occurred in the two days between 1919 December 4 and 6. Furthermore, there is uncertainty around the photometry since both photographic and visual data are combined in the lightcurve. In the absence of other data, we assume the maximum brightness was magnitude 6.5. Hence any value determined for decline parameters has an uncertainty of ±2 days. The fact that the HCO *Bulletin* reported the nova to be at 8.5 on 1920 Jan 6 suggests a possible value for $t_2$ of 31 days. However, we also fitted a sixth order polynomial to the data, between 1919 Dec 6 and 1921 Nov 24, which allowed us to determine $t_2$ = 47d. We note that the Jan 6 observation lies below our fitted line, which may suggest that the decline was initially faster. On the other hand the HCO *Bulletin*'s comment about 'marked fluctuations in brightness' in the period immediately after discovery leads us to question how representative this observation was. The basic problem is that there are insufficient observational data to be certain about the actual decline in this period, so we will proceed with both values of $t_2$ being equally plausible.

Our sixth order model gives $t_3$ = 97d, which makes HR Lyr a classical fast nova. Our $t_3$ value is slightly longer than the values of $t_3$ reported by Duerbeck, of 74 and 80d,[10,11] and by Allen, of 70d,[12] which were presumably based on a more limited dataset.

Duerbeck has also classified novae depending on the shape of the lightcurve.[10] Accordingly we assign HR Lyr to the type B, which he describes as 'decline with minor or major irregularities'. We consider there to have been some irregularities based on the comment in the discovery announcement about 'marked fluctuations in brightness' quoted above. Furthermore, we note considerable scatter, up to 1 magnitude, during the transition phase of the lightcurve, which may indicate further fluctuations. Duerbeck himself classified HR Lyr as 'possibly type A' in his 1981 paper,[10] but he gave no classification in his 1987 catalogue of galactic novae.[11] Further classification would require detailed spectroscopic analysis, but such analysis was limited during the outburst.[8]

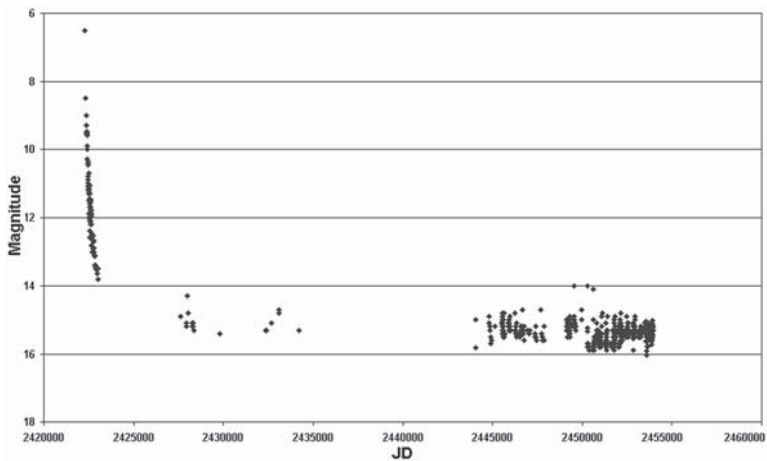

**Figure 2.** Lightcurve of HR Lyr from 1919 to 2006. Data are from sources listed in Figure 1, plus references 16 to 21 and the BAA VSS database.

## Maximum absolute magnitude

The decline parameters of novae have been linked to their maximum absolute magnitude and various formulae for this relationship have been developed. We have used our values of $t_2$ in three of these equations, each based on a different model,[13,14,15] to estimate the maximum absolute magnitude, $M_{(max)}$, and the results are shown in Table 1 (errors given include the errors in the formulae, plus uncertainties in the value of $t_2$). The mean values of $M_{(max)}$ of −6.9 and −7.2 (±1.1) are typical of a galactic nova.

## Return to quiescence

We have compiled as complete a lightcurve of HR Lyr as possible covering the period from the detection of the outburst up to 2006 July (Figure 2). This is based on more than 1000 positive observations obtained from the BAAVSS, AAVSO and AFOEV databases, as well as literature sources; 'less than' and 'uncertain' estimates have been omitted. Rather few positive observations exist between 1921 November (at which point the nova was still in decline, but was apparently becoming too faint, at 13.8v, for the visual observers of the time to follow) and 1979 July. During this time, most of the observations were contributed by W. H. Steavenson.[16–21] He commented that there was a small brightening in 1949 June, when the star rose from its normal magnitude of 15.1 to about 14.7, but by the following June (1950) it had returned to 15.3.

It is evident from the lightcurve that considerable scatter in the data appears to exist during quiescence, the origin of which will be discussed in subsequent sections. The average of all the observations made after 1979 July is 15.4 (SD is 0.23). However, we would point out that the accuracy of the visually observed magnitudes at minimum depends entirely on the accuracy of the comparison star sequence that the observers used. It appears that most observers have used the sequence published by Steavenson, based on estimates he made by eye.[17] By contrast, Kafka & Honeycutt have recently presented V-band photometry from RoboScope over a 9-year period which suggests that the average minimum magnitude of HR Lyr is closer to 16.0V.[24] This strongly suggests a zero-point error in the observations based on the Steavenson sequence. Furthermore, the comment in the HCO *Bulletin* about the star being '16th magnitude or fainter' before the outburst also points to a fainter quiescence. Hence we prefer to adopt a quiescence magnitude of 16.0 (rather than 15.4), which implies an outburst amplitude of at least 9.5 magnitudes. This is relatively modest for a nova outburst, which is typically in the range 8–15 magnitudes.

## Activity at quiescence

As mentioned in the previous section, there is a considerable spread in the archival observations made during quiescence. We have looked for periodicity within these data in the range 3 to 300d, using a variety of statistical tools, without success. Some scatter could be due to the inherent variations in variable star estimates (mostly visual in this case, but including some CCD) made by different observers, using different instruments and different comparison stars, to name but a few likely causes. However, the actual amount of scatter, which is more than 1 mag, is greater than we would expect and to some extent probably reflects real magnitude variations during quiescence.

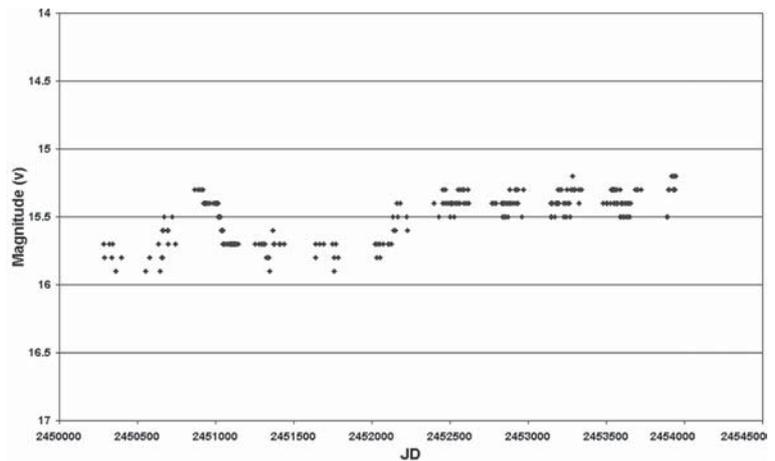

**Figure 3.** Visual observations of HR Lyr between 1996 July and 2006 July. *(Gary Poyner)*

In fact, there is much evidence in the literature that HR Lyr has an active quiescence, with variations on many timescales. Several people have reported inter-night brightness variations. For example, Bruch found HR Lyr at 15.82V on 1979 July 15, but 6 days later it had brightened to 15.00V; he also noted colour variations, since the B–V values on the two nights in question were +0.01 and +0.30 respectively, and U–B values were –1.06 and –0.60.[22] He found that brightness and colour variations are common in old novae, but in the case of HR Lyr they are particularly pronounced. In an extensive study conducted by a team at the Wise Observatory over 5 years in the 1990s, a 64d period variation of about 2 magnitudes in R was found.[23] They also found intra-night variations. These included rapid variations of several tenths of a magnitude over timescales of minutes and hours, which is probably 'flickering' – apparently random light variations with amplitudes of some tenths of a magnitude on a timescale of seconds or minutes – such as is associated with most cataclysmic variable stars.[3] Flickering has also been recorded by one of the present authors, JS, in unfiltered CCD photometry when variations of up to 0.1 mag have been detected over a 10-minute period.

The Wise Observatory team also found quasi-periodic variations around the period 0.1d, which they speculate may be associated with the orbital period. No independent measurement of the orbital period of HR Lyr has been published.

Kafka & Honeycutt's 9-year RoboScope data also showed variations between 15.2V and 16.6V,[24] although there was no evidence of a consistent periodicity.

**Table 1. Maximum absolute magnitude of HR Lyr**

| Method | $M_{(max)}$ based on $t_2$=31d | $M_{(max)}$ based on $t_2$=47d |
| --- | --- | --- |
| Linear model for galactic novae[13] | −7.1 ± 0.7 | −6.7 ± 0.7 |
| S-shaped model for extragalactic novae[14] | −7.4 ± 0.9 | −7.1 ± 0.8 |
| Linear model for novae of types B, C, D[15] | −7.2 ± 1.8 | −7.0 ± 1.7 |
| *Average* | **−7.2 ± 1.1** | **−6.9 ± 1.1** |

# Long term variations in the visual lightcurve during quiescence

One of the authors, GP, has followed HR Lyr since the mid-1990s. Since these observations have been made by a single, experienced, observer they have the advantage of being an internally consistent dataset. The resulting visual lightcurve comprising almost 300 observations (Figure 3) appears to show that the star remains for long periods at around 15.4v, whereas at other times it is fainter, at about 15.7v; each of these periods is joined by a transition during which the star gradually brightens or fades.

Although on a smaller scale, such behaviour is reminiscent of the 'high state' and 'low state' exhibited by several types of cataclysmic variable stars, especially the magnetic cataclysmic variables.[2] There are two classes of magnetic CVs: Polars and Intermediate Polars. Both classes contain representatives of old novae.[25] Polars have a very strong magnetic field of the order of 10–100 megaGauss: the field is so powerful that it prevents the formation of an accretion disc and locks the two stars into a synchronous rotation. V1500 Cyg is an old nova which is an example of a polar. In the case of intermediate polars, the magnetic field is less strong, of the order of 1–10 megaGauss, which allows an accretion disc to form although it is disrupted close to the primary. Furthermore, the magnetosphere is not strong enough to synchronise the rotation. Old novae that are examples of intermediate polars include DQ Her, GK Per and V533 Her.

However, the 0.3 magnitude changes which we observed in HR Lyr are rather modest compared to those which are generally found in polars. In a recent survey of fourteen polars, Ramsay found that the average difference between high and low states was 2.1 mag, although two stars, V393 Pav and CV Hyi, showed differences of 0.5 mag or less.[26] Intermediate polars also show high and low states, but this phenomenon appears to be less common than in polars.[27]

Further studies by professional astronomers, including spectroscopy, polarimetry and X-ray observations, could determine whether HR Lyr is indeed a magnetic star.

# Is HR Lyr a recurrent nova?

Although only one outburst of HR Lyr has been detected, over the years it has been speculated that HR Lyr may in fact be a recurrent nova. In 1967, Commission 27 of the IAU presented a list of 11 old novae as possible recurrent nova candidates, with the recommendation that these objects should be monitored for further outbursts.[28] These novae were selected because their outburst amplitude was relatively modest compared to most classical novae (HR Lyr is listed as having an amplitude of 8.5 mag, which according to our analysis above is probably an underestimate) and it was

**Table 2. Properties of HR Lyr and T Pyx**

*(Data for HR Lyr are from the current study, unless otherwise referenced)*

|  | HR Lyr | T Pyx |
|---|---|---|
| Outburst amplitude, mag | 9.5 | 8.3[30] |
| Time to maximum light, days | <2 | ~32[30] |
| $t_2$, d | 31 or 47 | 62[30] |
| $t_3$, d | 97 | 87[11] |
| $M_{(min)}$ | 2.3 or 2.6 (±1.1) | 2.4±1.6[29] |
| Orbital period, d | 0.1[23] | 0.076[33] |
| Recurrence time, years | – | ~19[30] |

thought that modest amplitude was linked with relatively short outburst intervals:[29] 'these are consequently objects that are likely with a certain degree of probability to undergo a further outburst in this century or shortly afterwards.' It is partly because of this call to arms that amateur astronomers have intensively monitored HR Lyr in recent years, but as yet no further outbursts have been detected.

There are currently six known recurrent novae, which have recurrence periods of ~20–80 years and outburst amplitudes of ~7–11 mag. In their definitive paper 'The Nature of Recurrent Novae',[29] Webbink *et al.* proposed two criteria which can be used to distinguish recurrent novae:

– the star must exhibit two or more outbursts, reaching an absolute magnitude at maximum comparable with those of classical novae (i.e. brighter than $Mv_{(max)}$ –5.5), and

– the ejection of a shell with an expansion velocity of at least 300km/s.

The first criterion distinguishes recurrent novae from both classical and dwarf novae and also from symbiotic stars; the second distinguishes them from the remaining symbiotic stars, many of which show bright, multiple outbursts, but without high-velocity shell ejection.

Webbink further proposed two classes of recurrent novae, based on their outburst mechanisms.[29,30] In one class, containing T CrB, V745 Sco and RS Oph, eruptions are powered by accretion events driven by a burst of mass transfer from a red giant secondary star.[31] In the other class, containing U Sco, V394 CrA and T Pyx, outbursts result from a thermonuclear runaway of accreted material on the surface of the white dwarf.

Are there any similarities between HR Lyr and the recurrent novae? Well, if HR Lyr is a recurrent nova, it is unlikely to be a member of the former class, involving a red giant secondary. The orbital period of such systems is of the order of several hundred days; RS Oph, for example, has a period of 230d.[30] As mentioned above, it has been proposed that the orbital period of HR Lyr is around 0.1d. Furthermore, the spectrum of such systems at quiescence is dominated by the secondary and hence these are red stars, whereas HR Lyr is a blue star at quiescence, with a continuum resembling a spectral class O star.[6–8] On the other hand, the proposed orbital period of HR Lyr is more similar to the members of the thermonuclear runaway class, whose periods range between 0.07 and 1.23d.[30] Hence if HR Lyr is indeed a recurrent nova it is more likely to be a member of the thermonuclear runaway class.

A characteristic of recurrent novae is a relatively bright absolute minimum magnitude, $M_{(min)}$. Webbink has set an upper limit on this value for members of the thermonuclear runaway class of $M_{(min)} \leq +3.6$, which is brighter than for most classical novae. In a study of 26 classical novae at minimum, the average $Mv_{(min)}$ was found to be +4.7, with a range between +1.90 to +6.20.[15]

We have estimated the value of $M_{(min)}$ for HR Lyr as either +2.3 or 2.6 (±1.1), based on our two average values for $M_{(max)}$ given in Table 1 and an outburst amplitude of 9.5 magnitudes. Both values of $M_{(min)}$ are within the limit set by Webbink for recurrent novae and are comparable to $Mv_{(min)} = +2.4$ (±1.6) for the recurrent nova T Pyx.[29] We compare some properties of HR Lyr and T Pyx in Table 2. In addition to similarities in $Mv_{(min)}$, they have rather similar values of $t_3$ and, possibly, orbital period. It should be noted that their $t_3$ values are much higher than those of the other members of the thermonuclear runaway recurrent novae, U Sco and V394 CrA, which are typically 5–6d.[32] The rise profile of T Pyx is very different in that it shows slow outbursts. In the 1966 outburst of T Pyx it took 32d to reach maximum light, whereas U Sco, V394 CrA and HR Lyr have each shown a very rapid rise, within 2 days.[29]

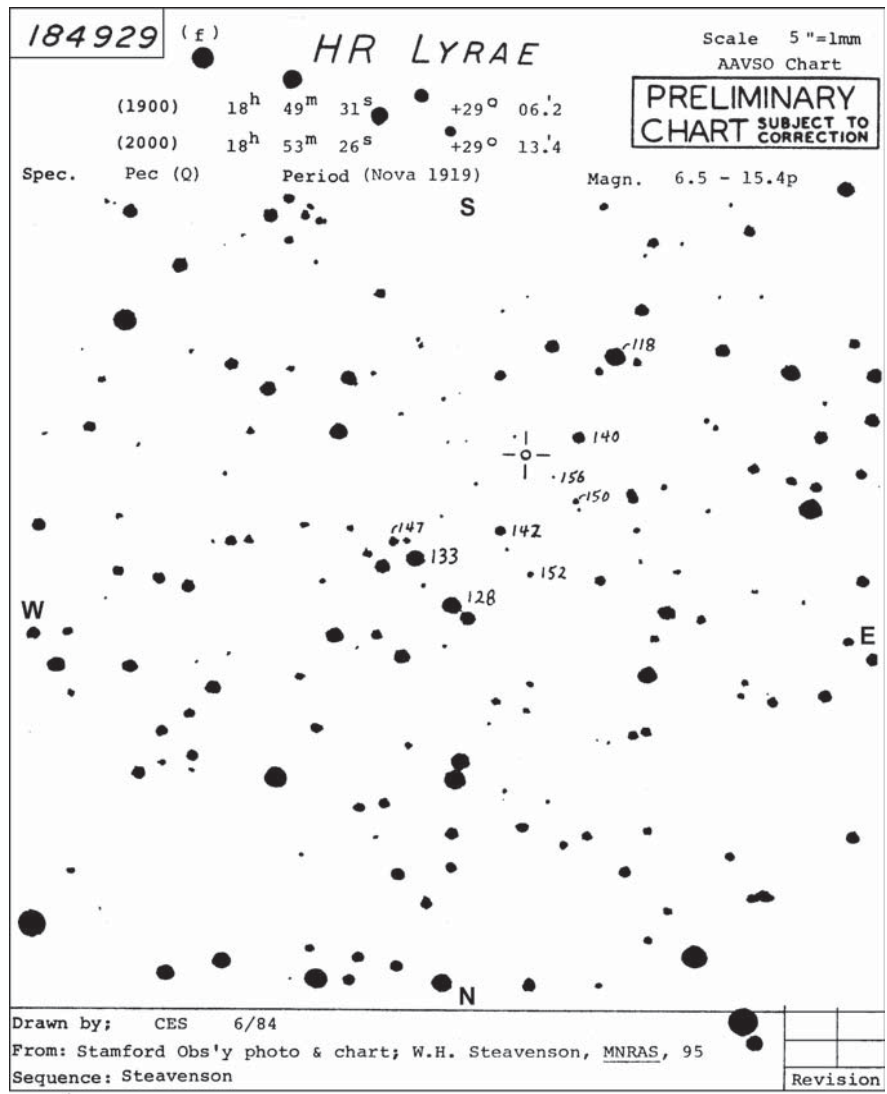

**Figure 4.** AAVSO finding chart for HR Lyr. Comparison star magnitudes have the decimal point omitted for clarity (e.g. star 140 is magnitude 14.0).

Thus there do appear to be some superficial similarities between HR Lyr and some of the recurrent novae, especially T Pyx. However, we must note that the average outburst period for T Pyx is 19 years, and yet it is more than 85 years since HR Lyr's entry onto the stage. Of course, whether HR Lyr itself is such an object can only remain speculation until another outburst is detected.

## HR Lyr and the Recurrent Objects Programme

The Recurrent Objects Programme (ROP) was set up as a joint project between the BAAVSS and *The Astronomer* magazine, specifically to monitor poorly studied eruptive stars of various types.[34] As a result of the speculation about HR Lyr being a recurrent nova, it was added to the ROP in 1993 by GP. We encourage all observers, both visual and CCD, to monitor this star for unusual activity. Whilst it would obviously be exciting to detect another outburst, it is also important to obtain further data about HR Lyr's active quiescence. As discussed earlier, HR Lyr's light varies on every possible timescale and important

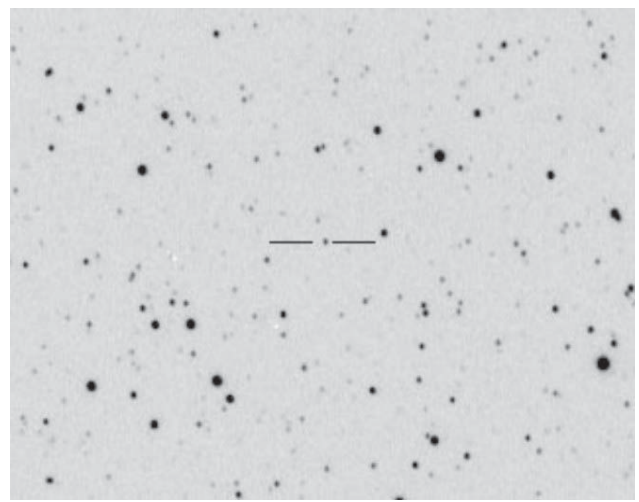

**Figure 5.** CCD image of the field of HR Lyr. Takahashi 0.1m apochromatic refractor, Starlight Xpress SXV-M7 CCD, 60sec integration (S is up, E to the right, field width 10'). *(Jeremy Shears)*

work can be done on refining these periods and possibly uncovering new ones. Given the reported colour changes, we suggest CCD observations are carried out using standard photometric filters. Charts for HR Lyr are available from the AAVSO.[35] The chart used by the authors is shown in Figure 4, along with a CCD image of the field (Figure 5) to aid identification.

## Summary

Based on our analysis of observations in the literature and in variable star databases, we have been able to produce a detailed lightcurve for HR Lyr and conclude this was a classical fast nova, probably of type B. Decline times were $t_2 = 31$ or 47 days and $t_3 = 97d$. The amplitude was 9.5 magnitudes. From the two possible values of $t_2$, we estimate that the maximum absolute magnitude was $-6.9$ or $-7.2$ ($\pm 1.1$) and the minimum absolute magnitude is $+2.3$ or $2.6$ ($\pm 1.1$).

The star appears to show an active quiescence with brightness variations on a variety of timescales. Visual observations conducted by one of the authors over a period of ten years also reveal long periods when the star is around 15.4v and others when it is around 15.7v.

We point out that some characteristics of HR Lyr are similar to those of recurrent novae. In particular, we note similarities between HR Lyr and the recurrent nova T Pyx in terms of outburst amplitude, absolute minimum magnitude, $t_3$ and, possibly, orbital period. We suggest that further monitoring of this star is warranted, not only to look for future outbursts, but also further to characterise its light variations at quiescence.

## Acknowledgments

The authors gratefully acknowledge the observations of HR Lyr contributed by observers worldwide to the databases of the BAAVSS, the AFOEV and the AAVSO which have been used in this research. We thank Dr Arne Henden, Director of the AAVSO, for giving his permission to reproduce the AAVSO chart of HR Lyr. We are indebted to Dr Chris Lloyd (Open University) and Guy Hurst (*The Astronomer*) for their helpful suggestions on the draft of this paper, to Professor Kent Honeycutt (Indiana University) for providing an insight into his long-term observations of HR Lyr in quiescence, and to the referees for constructive comments that improved the paper. Finally we thank John Toone and Roger Pickard (BAAVSS) for assisting in the search for archival observations of the star.

**Addresses: JS:** 'Pemberton', School Lane, Bunbury, Tarporley, Cheshire, CW6 9NR, UK [bunburyobservatory@hotmail.com]
**GP:** 67 Ellerton Road, Kingstanding, Birmingham, B44 0QE, UK [garypoyner@blueyonder.co.uk]

## References

1  Peltier L. C., *Starlight Nights: The Adventures of a Stargazer*, Sky Publishing Corporation (1999)
2  Hellier C., *Cataclysmic Variable Stars: How and why they vary*, Springer–Verlag (2001)
3  Bailey S. I., *Harvard College Observatory Bulletin* **705** (1920)
4  Grouiller M. H., *Journal des Observateurs*, **4**, 44–45 (1921)
5  Nijland A. A., *Bull. Astr. Institutes of the Netherlands*, **2**, 231–239 (1925)
6  Humason M. L., *Astrophys. J.*, **88**, 228–237 (1938)
7  Wyse A. B., *Publ. Astron. Soc. Pacific*, **52**, 334 (1940)
8  Wyse A. B., *Publ. Lick Observatory*, **14**, 229-233 (1940)
9  General Catalogue of Variable Stars: **http://www.sai.msu.su/groups/cluster/gcvs/gcvs/** (2006)
10  Duerbeck H. W., *PASP*, **93**, 165–175 (1981)
11  Duerbeck H. W., *Space Sci. Reviews*, **45**, 1–212 (1987)
12  Allen C. W., *Astrophysical Quantities*, 3rd edn, University of London, The Athlone Press (1973)
13  Cohen J. G., *ApJ.*, **292**, 90–103 (1985)
14  Della Valle M. & Livio M., *ApJ.*, **452**, 704–709 (1995)
15  Downes R. A. & Duerbeck H. W., *AJ.*, **120**, 2007–2037 (2000)
16  Steavenson W. H., *MNRAS*, **86**, 365 (1926)
17  Steavenson W. H., *MNRAS*, **95**, 639 (1935)
18  Steavenson W. H., *MNRAS*, **96**, 698 (1936)
19  Steavenson W. H., *MNRAS*, **97**, 655–656 (1937)
20  Steavenson W. H., *MNRAS*, **108**, 186 (1948)
21  Steavenson W. H., *MNRAS*, **113**, 258–261 (1953)
22  Bruch A., *IBVS*, 1805 (1980)
23  Leibowitz E. M. *et al.*, *Baltic Astronomy*, **4**, 453–466 (1995)
24  Kafka S. & Honeycutt R. K., *Rev. Mex. AA*, **20**, 238 (2004)
25  Cherepashchuk A. M. *et al.*, *Highly Evolved Binary Stars: Catalog* (Adv. in Astron. & Astrophys., **1**), Gordon & Breach (1996)
26  Ramsay G. *et al.*, *MNRAS*, **350**, 1373–1384 (2004)
27  Lipkin Y. M. *et al.*, *MNRAS*, **349**, 1323–1332 (2004)
28  Richter G. *et al.*, *Variable Stars*, Springer Verlag (1984)
29  Webbink R. F. *et al.*, *ApJ.*, **314**, 653–672 (1987)
30  Sekiguchi K., *Astrophys. and Space Sci.*, **230**, 75–82 (1995)
31  Schaefer B. E., *ApJ.*, **355**, L39–L42 (1990)
32  Sekiguchi K. *et al.*, *MNRAS*, 245, 28p–30p (1990)
33  Patterson J. *et al.*, *PASP*, **110**, 380 (1998)
34  Poyner G., *J. Brit. Astron. Assoc.*, **106**(3), 155–159 (1996)
35  **http://www.aavso.org/observing/charts**